# Ultra-High Density, High-Performance and Energy-Efficient All Spin Logic

Mrigank Sharad, Karthik Yogendra, Arun Gaud, Kon-Woo Kwon, Kaushik Roy, *Fellow, IEEE*
Department of Electrical and Computer Engineering, Purdue University, West Lafayette, IN, USA
(msharad, kyogengra, arun@purdue.edu, kon-woo.kwon.1, kaushik)@purdue.edu

*Abstract:* All Spin Logic (ASL) gates employ multiple nano-magnets interacting through spin-torque using non-magnetic channels. Compactness, non-volatility and ultra-low voltage operation are some of the attractive features of ASL, while, low switching-speed (of nano-magnets as compared to CMOS gates) and static-power dissipation can be identified as the major bottlenecks. In this work we explore design techniques that leverage the specific device characteristics of ASL to overcome the inefficiencies and to enhance the merits of this technology, for a given set of device parameters. We exploit the non-volatility of nano-magnets to model fully-pipelined ASL that can achieve higher performance. Clocking of power-supply in pipelined ASL would require CMOS transistors that may consume significantly large voltage headroom and area, as compared to the nano-magnets. We show that the use of 'leaky' transistors can significantly mitigate such bottlenecks, without sacrificing energy-efficiency and robustness. Exploiting the inherent isolation between the biasing charge-current and spin-current paths in ASL, we propose to stack multiple ASL metal layers, leading to ultra-high-density and energy-efficient 3-D computation blocks. Results for the design of an FIR filter show that ASL can achieve performance and power consumption comparable to CMOS, while the ultra-high-density of ASL can be projected as its main advantage over CMOS.

## 1. Introduction

Two different methods of current induced Spin-Transfer-Torque (STT) switching of *nano-magnets* have been demonstrated in recent years. The first involves injection of spin polarized charge-current (injected by another magnet) into a *nano-magnet*. The second strategy, called non-local-STT, employs pure spin-current injection into a *nano-magnet* [1], [2].

Fig. 1a shows the lateral spin valve (LSV) structure with non-local spin current injection. It consists of an injecting magnet ($m_1$) and a receiving magnet ($m_2$) connected through a non-magnetic channel. Electrons flowing into the channel through $m_1$ get right-spin polarized when they reach the magnet-channel interface. Spin-polarized charge-current is modeled as a four-component quantity, one charge component ($I_c$) and three spin components ($Is_x$, $Is_y$, $Is_z$) [3-5]. Due to the entire-channel being at the same charge-potential, there is no charge-current flow along it. All the charge-current flows into the ground lead, towards a lower charge potential. The left-spin magnet $m_1$ however imparts an excess of left-spin polarity to the electrons injected into the channel, creating a spin-potential difference across the channel. This results in a spin-diffusion current flow along the channel. The magnet-channel interface of $m_2$ can absorb the transverse components of the spin-current which in turn exerts spin torque on the receiving magnet and causes it to flip. Owing to the separation of the spin-diffusion current responsible for *nano-magnet* switching, from the charge current flow, such a spin transport in the lateral spin valve is termed as 'non-local'.

Analog characteristics of current mode switching employed in LSV's can facilitate non-Boolean computation like weighted majority evaluation. Hence, LSV's with multiple input magnets can be used to design spin majority gates. In [3] authors proposed 'All Spin Logic' (ASL) scheme that employed cascaded LSV's interacting through unidirectional, non-local spin current [4-14]. Unidirectionality is introduced by incorporating a high-polarization (high-*P*) interface for the transmitting side of a magnet and a low-*P* interface for the receiving side of a magnet as shown in fig. 1b. A high-*P* side injects strongly spin-polarized current, where as a low-*P* side absorbs spin-current [3]. Fig 1c and fig. 1d depict ASL NAND gate [3], and, ASL full-adder [6], based on spin-majority evaluation.

With a constant power supply, the switching speed of a large ASL circuit would be limited by the longest chain of nano-magnets present in it. Maximum performance for a large ASL block can be achieved by the use of two-phase pipelining. Since a *nano-magnet* preserves its state upon removal of supply voltage, ASL can facilitate fine-grained pipelining, without the need of additional latches. However, this comes at the cost of power and area overhead, resulting from the clocking transistors that are used to turn the supply on and off for a given logic stage. We show that this overhead can be significantly reduced by employing transistors with reduced on/off ratio. The non-local STT employed in ASL decouples the biasing charge-current path from the spin-current path along which computation is performed. Common charge current channels can therefore be shared by multiple ASL layers in a vertical stack, leading to a high-density 3-D spin logic design. In such a design, a single CMOS substrate can be used to supply 2-phase clock to a large number of low-resistance, metallic ASL stacks, leading to high energy efficiency along with very high integration-density [22]. Based on the simulation-based analysis, we show that the proposed design techniques achieve more than 2 orders of magnitude lower energy dissipation per computation, as compared to the standard ASL configuration [3].

In this work we have assumed optimal device parameters, like, high spin-injection efficiency, low contact resistances and efficient control over polarization constant (*P*) of magnet-channel interface and sufficient spin-diffusion length for non-magnetic channels. Techniques for improving these parameters are still under research. The target of this work however, is to project the best achievable performance for ASL, assuming that future research can achieve the near optimal device and material parameters.

Rest of the paper is organized as follows. Section 2 describes the device operation for 2-phase pipelined ASL and the associated device-level optimizations. Section 3 presents the design of a generic ASL block (like an 8-bit, 8-stage FIR filter) using 2-phase pipelining, along with the techniques for optimal clocking and 3-D integration and their associated benefits. Simulation-framework used in this work is described in section 4. Finally, section 5 concludes the paper.

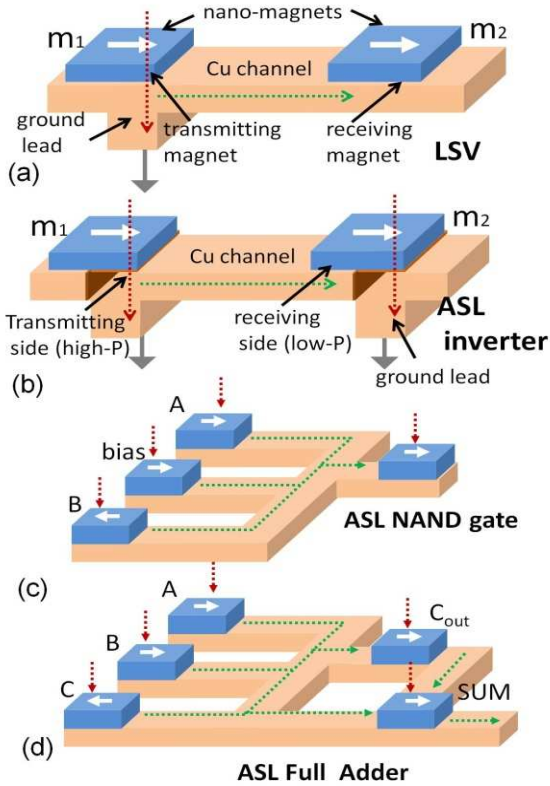

Figure 1(a) Lateral spin valve (LSV) demonstrated in [1], (b) ASL inverter [3](c) ASL NAND gate [3] (d) compact ASL Full adder using five magnets [6]. (e) Simulation waveforms for FA evaluation, (blue curve denotes the easy-axis magnetization component mz, mx, my being the orthogonal components

## 2. Two-Phase Pipelined ASL

In this section operation of 2-phase pipelined ASL is described. Following this a brief discussion on device level optimization is given.

*A. Device Operation*

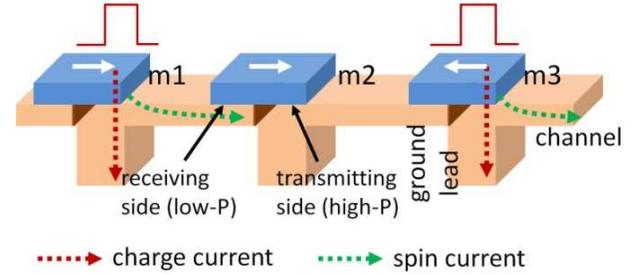

Fig. 2 Three ASL stages connected using 2-phase pipelined scheme

Fig. 2 shows three ASL stage connected using 2-phase pipelining scheme. The magnets $m_1$ and $m_3$ are driven by clock, where as $m_2$ is driven by an inverted clock. When the clock is high, $m_1$ and $m_3$ act as transmitting magnets. They receive charge current from a clocked transistor (not shown in the figure). The injected charge current induces a spin-diffusion current on the transmitting (high-P) side of $m_1$ and $m_3$, which in turn, is absorbed by the receiving magnets' low-P side (low-P side of $m_2$ as shown in the figure). Thus, the data stored in $m_1$ is transferred to $m_2$ and that stored in $m_3$ is transferred to the next magnet (not shown in the figure). During this phase, $m_2$ is not connected to the supply and hence does not receive any charge current. When the clock goes low, the magnets $m_1$ and $m_3$ turn into receivers and are kept in the floating state. For $m_3$, $m_2$ acts as the transmitter. Thus, the information stored in $m_2$ during the high-clock phase is transferred to $m_3$ during the low clock phase.

The same scheme can be extended to an arbitrary pipelined logic design. Fig. 3 shows two ASL full adders (FA) connected using the pipelining scheme described above. Here, pipelining-granularity has been taken as a single FA. Note that a single ASL-FA evaluation corresponds to two magnet switching delays. Current is supplied simultaneously to all the five magnets in a FA. First $C_{out}$ evaluates, based on the values of the inputs $A$, $B$ and $C_{in}$. This is followed by the evaluation of $SUM$, which depends upon the state of $C_{out}$ and the three inputs. In this example, the clock pulse width must be at least as wide as two magnet switching delays. A finer pipe-line granularity would involve decomposing one FA evaluation two steps, namely, $C_{out}$ evaluation and $SUM$ evaluation.

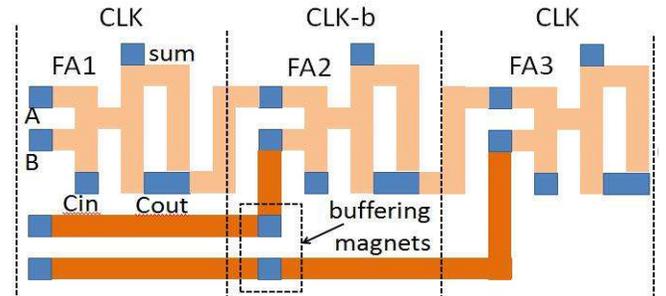

Fig. 3 ASL full adders connected using 2-phase pipelining scheme.

Note that, in contrast to pipelines CMOS, a 2-phase pipelined ASL does not require any additional latch, as each of the *nano-magnets* itself acts as a latch and hence, facilitates fine-grained pipelining of a large logic block.

## B. Device Optimization

The device operation for 2-phase pipelining can be optimized by appropriate choice of device structure and operating conditions. Fig. 4 depicts the spin-diffusion model for ASL device. The device elements, namely, the metal channel, the *nano-magnets* and the magnet-channel interface are modeled as four component conductance elements, one charge conductance and three spin conductance's ($G_{se}$, $G_{sh}$: series and shunt conductance of metal channel, $G_{int}$: magnet-interface conductance) [4, 5]. As mentioned earlier, the charge component of the spin-polarized current injected into the channel through the high-$P$ side of the transmitter magnet passes into the ground lead. In the pipelined scheme, since the receiving magnet is in floating state, there is no charge current flow in the channel and through the receiving magnet. A part of the spin component of the input current is also lost to the ground, whereas, the rest is absorbed by the receiving magnet. Increasing the length of the ground lead increases its charge resistance, as well as, its spin resistance. Hence, for a given input current $I_{charge}$, the spin current $I_{spin}$, absorbed by the receiving magnet increases with increasing ground resistance $R_g$ (fig. 5). The ratio of $I_{spin}$ and $I_{charge}$ can be defined as the non-local spin-injection efficiency (NLSE). Note that, experimentally ~20% efficiency for non-local spin injection has been demonstrated (with $P$~0.5) [1, 4]. In this work we used 25% NLSE in simulations.

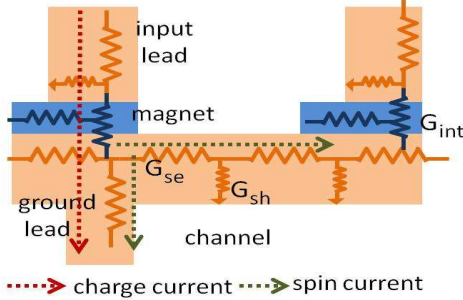

Fig 4 Spin diffusion model for an ASL device showing an input magnet and an output magnet connected using a metal channel.

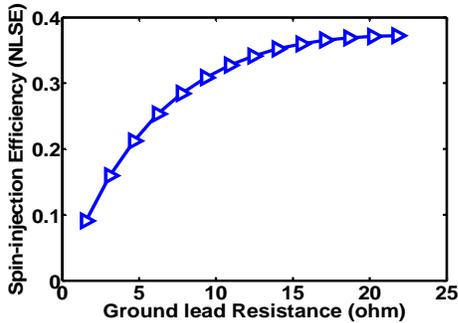

Fig. 5 Plot showing increase in non-local spin injection efficiency with increasing ground resistance.

Scalability of *nano-magnets* in ASL is also tightly coupled to the value of $R_g$. For a given input current, scaling down the area of a receiving magnet lowers its conductive interface with the metal channel, thereby resulting in lower spin-current absorption. However, as the density of spin-current absorption remains constant, a constant switching speed is maintained, as shown in fig. 6.

Benefit of *nano-magnet* scaling however, can be obtained by simultaneously scaling of $R_g$. As explained above, by reducing $R_g$ along with the magnet area, the spin injection efficiency is maintained. Hence, for a given charge current input, the value of spin injection is enhanced, leading to faster switching (fig 7). The ultimate scaling of *nano-magnets* in an ASL device will be therefore governed by the scalability of the ground lead (or via).

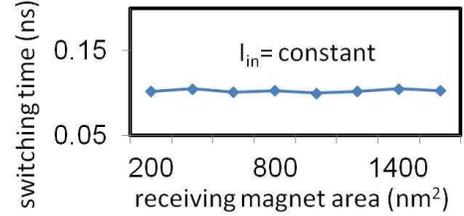

Fig. 6 Scaling of magnet area maintains a constant switching speed for a given input current and a fixed $R_g$

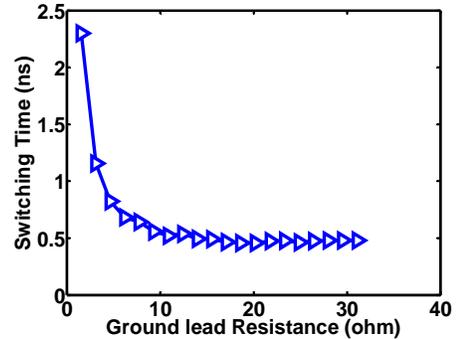

Fig. 7 Up-scaling of $R_g$ with reducing magnet area leads to faster switching

An important consideration for a pipelined ASL design is the choice of clock period. For a *nano-magnet*, switching energy, $E_{sw}$, can be expressed as $E_{sw} = T_{sw} \times I_{sw} \times V$, where $T_{sw}$ is the switching time and $V$ is the terminal voltage. To the first order, *nano-magnet* switching time is inversely proportional to the switching current (fig. 8). Since, higher $I_{sw}$ requires higher $V$, faster switching speed incurs linearly higher switching energy, as shown in fig. 9. Thus, for low-energy operation it is desirable to operate the pipelined ASL with a low frequency clock. However, in presence of thermal noise, the probability of correct switching of *nano-magnets* reduces steeply with reducing current.

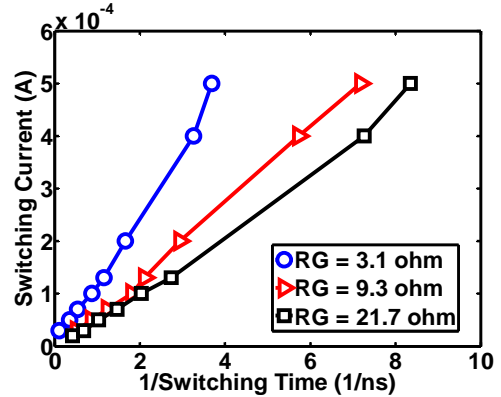

Fig. 8 Nano-magnet switching current increases linearly with switching frequency

Fig. 10 shows the plot for the probability of correct evaluation vs. $I_{sw}/I_{cr}$, where $I_{cr}$ is the critical current required to switch the *nano-magnet* in a long enough time. Stochastic-LLG has been employed to determine this trend [3]. This observation implies that the impact of thermal noise upon the bit-error rate may play a critical role in determining the lower limit of switching energy achievable for pipelined ASL.

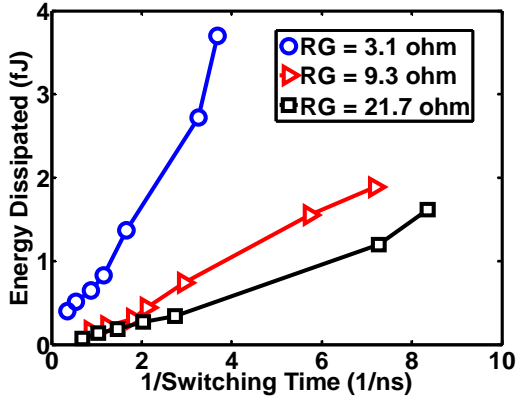

Fig. 9 Switching energy for ASL device increases linearly with switching speed.

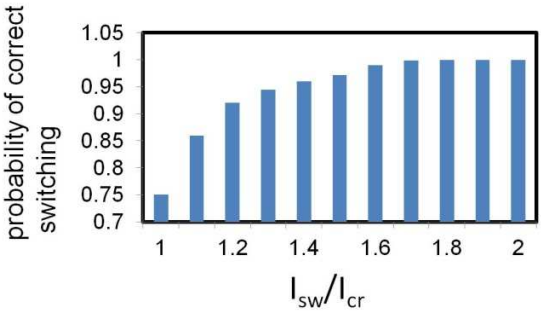

Fig. 10 Switching probability vs. $I_{sw}/I_{cr}$

## 3. Design-Optimizations for Pipelined ASL Circuits

In this section we present circuit-level design optimizations for pipelined ASL with the example of an 8-bit, 8-stage FIR filter (fig. 11). First, the layout and interconnect-design for such large ASL blocks are discussed. This is followed by optimization of the clocking transistors. Finally, techniques for achieving higher performance with parallelism and 3-D integration is described.

### 3.1 Layout and interconnects for ASL

The block diagram for a fully pipelined FIR-filter is shown in fig.11. The two main constituents of the filter; the multipliers and the adders can be pipelined with the finest granularity [18. 19]. These blocks mostly employ near-neighbor connectivity, as shown in the corresponding block diagrams in fig. 12. The corresponding ASL layouts are shown in fig. 13. The layout has been done using just two metal layers. For modular circuit units like arithmetic blocks, copper-channels with a spin-diffusion length ~1µm can be used for short distance links.

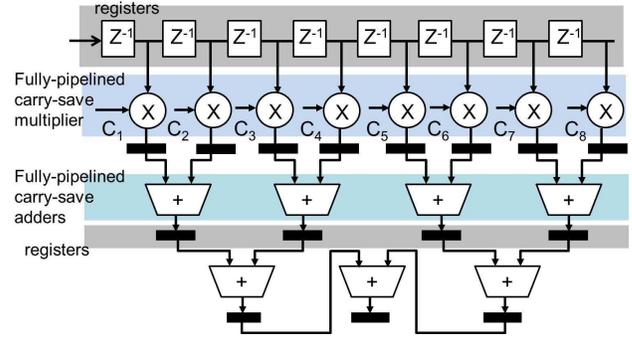

Fig. 11 Block diagram for a fully pipelined 8-stage FIR filter.

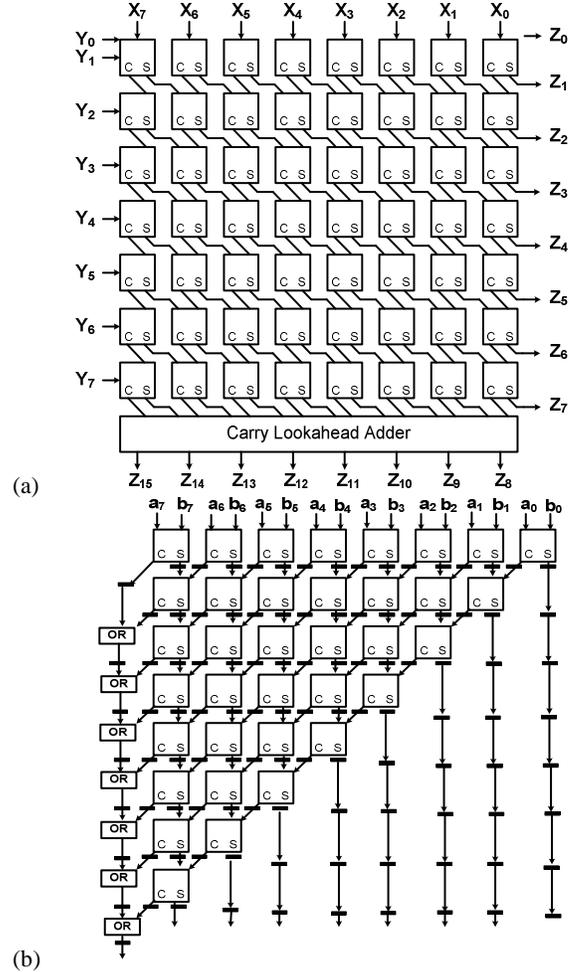

Fig. 12 Block diagrams for (a) a fully pipelined carry save multiplier, (b) fully pipelined carry save adder

However, for the integration of smaller blocks into larger designs, the complexity and routing distance of channels increase significantly. For longer distances, graphene channels can be employed [16]. Recent experiments have demonstrated large spin-diffusion lengths (distance after which spin-current decays by a factor of *e*) for graphene (~10µm). Techniques to improve spin-injection efficiency and contact resistance for graphene and nano-magnet interface have also been recently reported [17]. A 2-level interconnection scheme based on graphene is shown in fig. 14. At the transmitting side, it constitutes of an input magnet $m_1$ which injects spin-polarized current into the vertical Cu-via and induces 'local-STT' switching of the nano-magnet $m_2$. This is followed by a long-distance spin-diffusion transport along the graphene channel, resulting in non-local STT switching of $m_3$. Finally, at the receiving end, $m_3$ affects local-STT switching of $m_4$, through the Cu-via. In this case, the Cu-via play the dual role of supply rail as well as spin-channel. Such a possibility can be useful for 3-D integration of ASL as discussed in section 3.3. Also, note that the direction of currents in the Cu-vias at the transmitting and the receiving side need to be opposite, in order to realize a non-inverting transmission channel. This can be achieved by appropriate connections at the CMOS clocking interface discussed later.

The adder-tree employed in the FIR filter design, using the 2-level interconnection scheme discussed above, is shown in fig. 15. Channel-material in both the levels ($L_1$ and $L_2$) can be chosen to be graphene. Alternately, a combination of Cu-channel for short distance $L_1$ and graphene channel for long distance $L_2$ interconnects may also be employed.

Next, we discuss the optimization of clocking transistors needed to realize 2-phase pipelining in ASL.

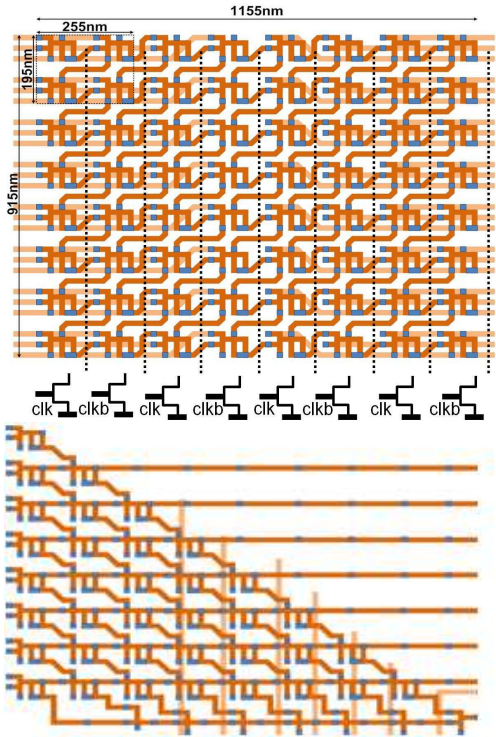

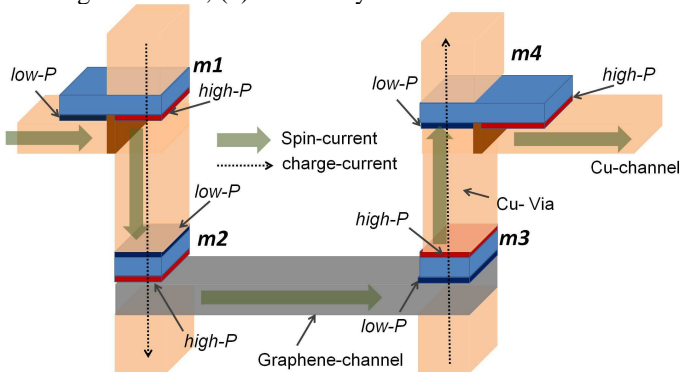

Fig. 13 ASL layouts for (a) 8x8 carry save multiplier with clocking transistors, (b) 8-bit carry save adder

Fig. 14 2-level interconnect scheme using long-distance graphene channel

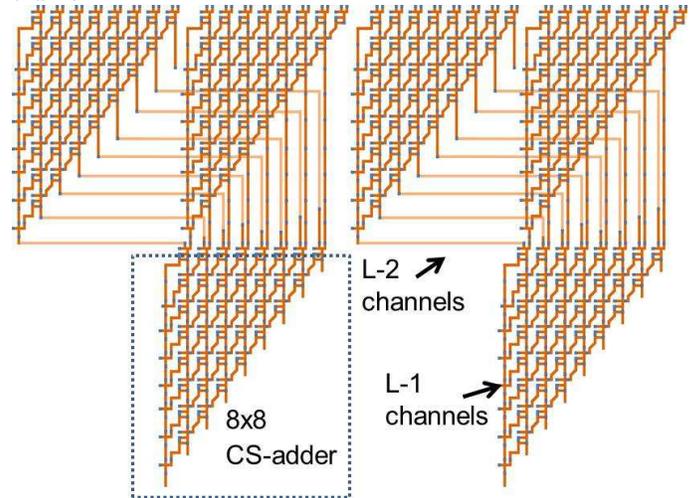

Fig. 15 Block diagram for 8-bit Carry-save multiplier with 2-levels of interconnections.

### 3.2 Optimization of clocking Transistors

*A. Trade-off between area and power:* The energy efficiency of pipelined ASL is strongly dependent upon the characteristics of the transistors used for clocking. In this work we chose full-adder delay as the single stage-delay of the pipeline. Each pipeline stage therefore constitutes of a parallel bank of full adders, as shown in fig. 13a (and 'buffers-magnets' in some paths for equalizing the path delays). Each such stage receives the bias current from a clocked CMOS transistor. The transistors belonging to the alternate stages are driven by complementary clock-phases (clk and clkb) in order to implement 2-phase pipelining (fig.13a). Owing to comparatively large resistance of the transistors, for a given switching delay, the drain-to-source voltage required is significantly larger than the voltage directly applied to the nano-magnets in the non-pipelined case. The voltage required can be reduced by increasing the width of the clocked transistors. But this leads to reduction in area-efficiency.

Minimum area for the ASL block is obtained when the area of the clocked transistors (Area_Tx) equals that of the ASL array (Area_spin). For instance, the area for an 8-bit ASL multiplier was estimated to be ~50x lower than that of a 15nm CMOS design. But for the minimum area case, the drain to source voltage required was found to be around ~140mV for 500MHz operation, and resulted in ~10X higher power consumption as compared to CMOS at iso-performance. The supply-voltage and hence, the power-consumption can be reduced by increasing the size (number of fingers) of the clocked transistors (fig. 16a). Note that, scaling down the supply voltage and scaling up the transistor widths by the same factor maintains the level of supply current per gate. As a result, static power involved in ASL computation is lowered (fig .16b). However, due to increase in the width of the clocked transistors, the dynamic clocking power increases (fig. 16b).

Fig. 16c shows the plot for area saving obtained by the ASL multiplier over 15nm CMOS multiplier for increasing transistor width. The corresponding trend for power consumption (relative to CMOS) is shown in fig. 16d.

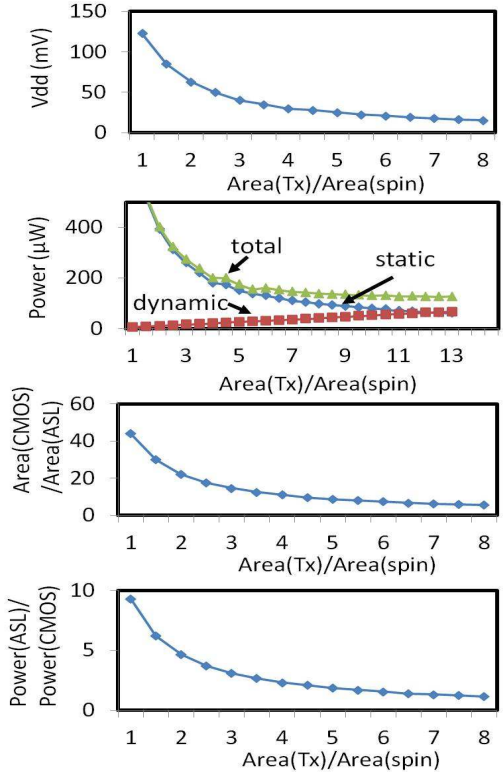

Fig. 16 (a) Supply voltage needed for pipelined ASL (for 300MHz operation) reduces with increasing Area(Tx), (b) Higher transistor width and lower supply voltage leads to reduction in static power but the dynamic clocking power increases. (c) Area benefit of pipelined ASL over 15nm CMOS design reduces with increasing Tx area (as expected). (d) Power consumption of pipelined ASL reduces with increasing Tx area and reaches a minima, but saturates after a certain point due to increase in dynamic power consumption.

From the foregoing discussion, it is apparent that in the pipelined ASL scheme, application of standard CMOS transistors for clocking leads to stringent tradeoff between power and area-efficiency, thereby eschewing the overall benefits of 2-phase clocking. However, despite achieving poor energy-efficiency, the minimum area ASL using standard CMOS transistors provides advantage in terms of robustness. For the non-pipelined design, the operating speed is determined by the critical delay path in the multiplier block. Hence, for achieving high frequency operation, current per magnet must be increased. Simulation results show that a non-pipelined 8x8 multiplier requires ~10x higher current injection to achieve the same performance as the pipelined design. However, since there are no transistors, the voltage required is also significantly low. As a result, non-pipelined case consumes only ~1.5X higher power than the minimum area pipelined design. But, the main bottleneck of the non-pipelined design is the high current requirement, leading to unreliably high current density (~$10^8$ A/cm$^2$) in the metal leads.

Next, we describe the application of 'leaky' transistors (transistors with low on-off current ratio) for clocking that can overcome the aforementioned design bottleneck for pipelining, and can simultaneously achieve low voltage operation and high area density.

### B. 'Leaky' transistors for designing energy-efficient pipelined ASL

Employing larger transistor area for obtaining higher on-current for the ASL gates leads to increase in dynamic power and area. An alternate method of reducing the on-resistance of the clocking transistors (and hence the required supply voltage) can be the use of excessively high source-drain doping. This technique can reduce the on-current of a transistor by an order of magnitude (fig. 17), thereby facilitating the use of smaller transistor for a given current-requirement and voltage supply. However, this technique leads to a much larger degradation in the $I_{on}/I_{off}$ ratio. The increase in off-current can be more than an order of magnitude higher than that in the on-current. However, the increase in average static-current can still be small. For instance, even with an $I_{on}/I_{off}$ ratio of ~10, the current flow during the off-cycle is only ~10% of that during the on period, implying only ~5% increase in the average static-current flow. On the other hand, the reduction in the on-resistance and hence in the drain-to-source voltage of the transistors can be larger than a factor of ~10, leading to commensurate reduction in static-power dissipation. The reduction in transistor switched capacitance simultaneously provides proportional reduction in the dynamic power.

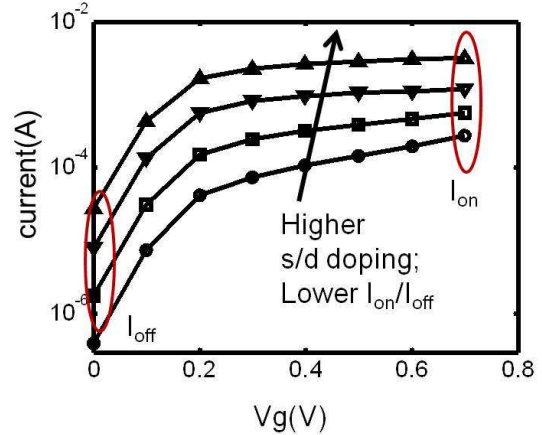

Fig. 17 Id-Vgs characteristics of 15nm NMOS for different $I_{on}/I_{off}$ ratio

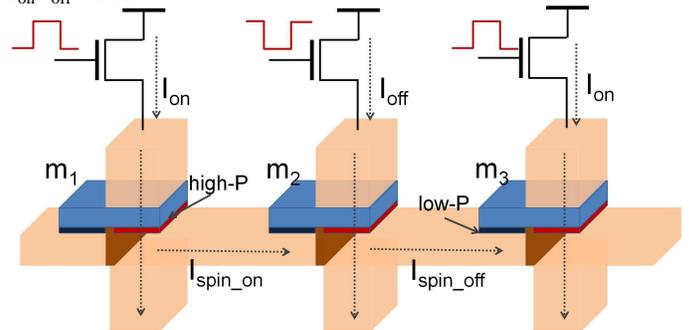

Fig. 18 Three pipeline stages of ASL

An important concern with the use of transistors with poor $I_{on}/I_{off}$ ratio can be the robustness of the logic operations. Fig. 18 depicts an exemplary ASL pipeline with one magnet per-stage. The off-state leakage-current ($I_{off}$) injected into the channel through the high-*P* side of a receiving magnet ($m_2$) can disturb the state of a transmitting magnet ($m_3$) in the next stage. Hence a safe limit for $I_{on}/I_{off}$ ratio must maintained.

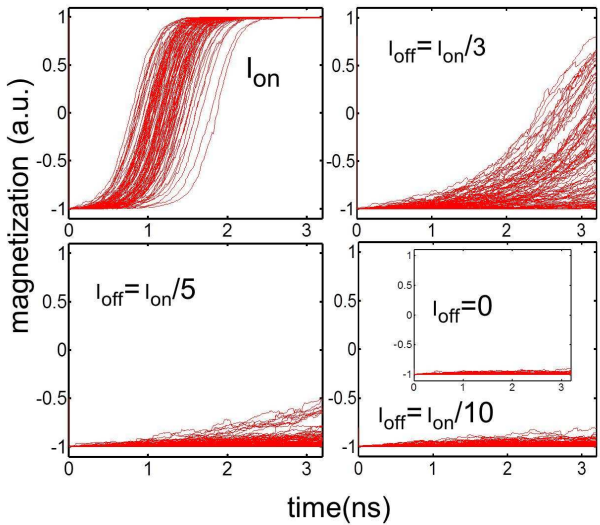

Fig. 19 Stochastic LLG simulation plots for magnet dynamics under the application of *ON*-current and different levels of- *OFF* currents.

The switching speed of a nano-magnet increases linearly with switching current [5]. Hence, a factor of ~10X reduction in current can be expected to increase the switching delay by the same amount, thereby providing a sufficient disturb-margin. Fig. 19 shows the simulation of nano-magnet dynamics under the application of *ON*-current (corresponding to 3ns switching time) and three different levels of *OFF*-current. Stochastic Landau-Lifshitz –Gilbert (LLG) equations have been used to capture the effect of thermal noise [3]. The plots show that an $I_{on}/I_{off}$ ratio of ~10 may provide sufficient robustness for error-free logic flow in pipelined ASL. Note that a much lower read-disturb margin (read to write current ratio of 2 to 3) is commonly used in magnetic random access memory (MRAM) [21]. However, in ASL heating effects can be more prominent due to continuous injection of bias currents.

In the present work the 8x8 multiplier can be identified as the densest in terms of the number of nano-magnets per unit area. For such a block, the application of the foregoing optimization facilitated the use of minimum transistor area (for which ASL area equals clocked transistor area) with 15mV supply ($V_{ds}$). Nominal transistors would consume ~10X larger area for the same $V_{ds}$. In the following section, we present a 3-D integration scheme that can further enhance the energy and area-efficiency of ASL.

### 3.3 3-D ASL for Ultra High Density and Low Power Computation Blocks

The inherent isolation between the bias current-path and the spin-current path in ASL can be exploited for 3-D integration. The proposed 3-D ASL design, shown in fig. 20, constitutes of multiple ASL layers stacked vertically. In such a design, each horizontal, 2-D layer performs computation independently. *Nano-magnets* in multiple 2-D layers sharing the same 2-D coordinates can be supplied charge-current through a common via. A spin-scattering layer can be deposited on the top of each magnet in order to prevent spin current interaction along the vertical vias.

For the 3-D design, the same transistor can supply current to all the magnets connected to a particular set of vias in all different layers. Thus, the power dissipation as well as the area of the clocking transistors is shared by all the ASL layers in the vertical stacks. Since, the overall resistance of the metallic vias is small as compared to the transistors, there is only a small increase in the supply voltage as compared to the one-layer, pipelined-ASL, for a given current-level (fig. 21c). This implies that the increase in the total power consumption remains small. Hence, the effective power consumption per ASL layer reduces, as long as, the *ON*-resistance of the transistors dominate the via resistance of the multiple stacks. The effective energy-efficiency as well as the area benefit over CMOS is therefore, significantly enhanced. Fig. 21 shows that for minimum transistor-area (using leaky transistors), stacking of 8 ASL layers can achieve ~6X reduction in effective power consumption. The main advantage of 3-D stacking however can be visualized in terms of area efficiency. Since the same CMOS substrate-area is employed for supporting multiple magneto-metallic ASL layers, the area density of ASL is enhanced by the number of ASL layers stacked.

In the proposed scheme, the clocked charge-current injection occurs along the vias in the vertical direction, whereas, the spin diffusion current in each individual layer flows along the metal/graphene channels along the horizontal direction. Spin-mode communication can also be established among adjacent 2-D layers through spin-polarized current injection along the vias, as discussed earlier. In the next section we summarize the performance of the design-techniques for ASL proposed in this work.

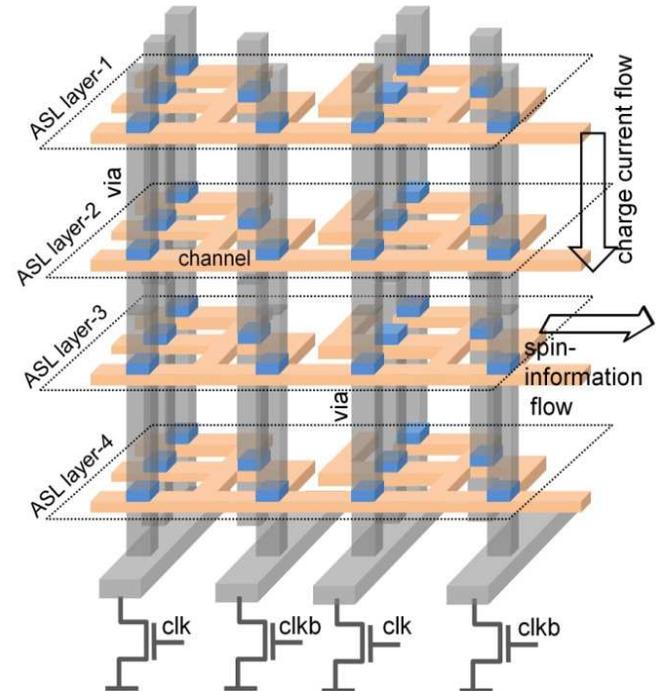

Fig. 20 3-D ASL can be constructed by stacking multiple ASL layers along the vertical direction. All the ASL layers in the vertical direction are supplied current using the same CMOS transistors.

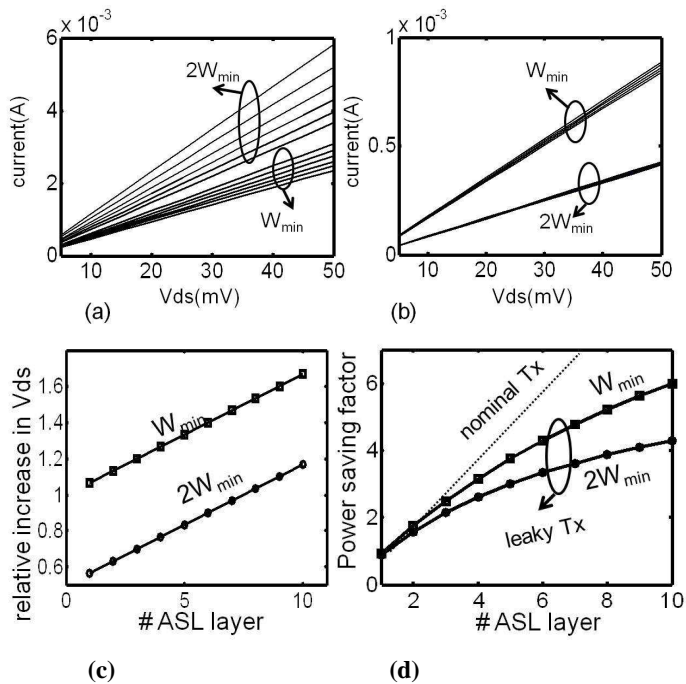

Fig. 21 (a) Due to low resistance of the metallic vias, there is only a small reduction in bias current for a given transistor width and supply voltage, even when multiple ASL layers are stacked, (b) this change in the case of nominal transistors is negligible due to higher *ON*-resistance, (c) relative increase supply voltage for a constant current-level with increasing number of ASL stacks. $W_{min}$ denotes the minimum area case, (d) power saving as a function of number of ASL stacks.

## 4. Performance Summary

Fig. 22 Depicts the benefits of the design techniques presented in this work. As mentioned before, the key feature of ASL is its compactness, whereas, the energy-inefficiency resulting from relatively larger magnet-switching delay can be identified as the down-side of it. In a non-pipelined ASL design, achieving 500MHz operation for an 8x8 multiplier would mandate ~60ps switching-speed for individual magnets (clock time period divided by total number of magnets in the critical logic path of the multiplier), requiring untenably high current-levels for magnets. This would result in large static-power in the ASL device (which is the only power component for a non-clocked, non-pipelined ASL) as shown in the figure. Introduction of minimum area clocking-transistors for 2-phase pipelining can help reduce the magnet-switching time (and hence the current-levels), leading to large reduction in power-dissipation in the device. As mentioned before, the switching speed of a nano-magnet is proportional to the switching-current. Hence, the $I^2R$ power-dissipation ($I$ = current, $R$ = device resistance) in the ASL device bears a quadratic dependence upon the switching-speed. The use of relatively high-resistance transistors however requires significantly higher voltage (~100mV as compared ~10mV), and leads to significant static-power dissipation across the transistors. Such a minimum area design barely offers any energy benefits over non-pipelined ASL, but does improve the reliability of the design by reducing the current injection for a given circuit performance by more than an order of magnitude. Area can be traded-off with power by increasing the size of the clocking transistors. This reduces the associated $I^2R$ dissipation in the transistors. However, this results in increase in dynamic power consumption due to increased switched capacitance. A more effective way of simultaneously achieving low-area as well as energy-efficiency can be the use of leaky transistors for clocking ASL. Such transistors with relatively low on-resistance achieve reduced static power dissipation and can simultaneously minimize the dynamic switching power by facilitating the use of smaller transistor sizes for a given current level. Finally, we proposed the design of 3-D ASL that can further enhance both energy-efficiency and area-density, by sharing a common CMOS substrate with multiple ASL layers.

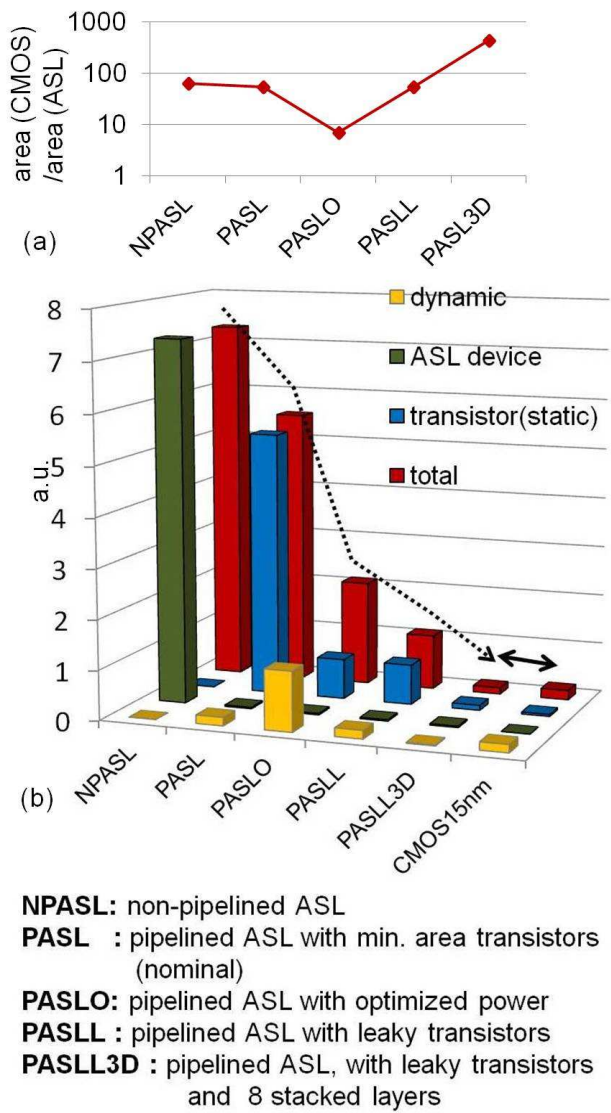

**NPASL:** non-pipelined ASL
**PASL** : pipelined ASL with min. area transistors (nominal)
**PASLO:** pipelined ASL with optimized power
**PASLL** : pipelined ASL with leaky transistors
**PASLL3D** : pipelined ASL, with leaky transistors and 8 stacked layers

Fig. 22 (a) Area comprison of proposed ASL design with CMOS, showing the possiblity of 3 order of magnitude higher logic density, (b) different components of power dissipation in the proposed scheme and their comparison with standard ASL and CMOS.

Results for 8-bit, 8-stage FIR filter design show the proposed design techniques can lead to ~60x lower power consumption as compared to a non-pipelined ASL for 500MHz throughput. For a given performance, the energy-benefit of pipelined 3D-ASL increases with the complexity of the design. This is because, performance of a pipelined ASL design is independent of the logic depth (or critical delay). On the other hand, for a given supply-voltage, the performance of a non-pipelined ASL would reduce with increasing critical-path-delay. Notably, results show that the proposed design-scheme can achieve power consumption, comparable to CMOS, provided, desirable device parameters (like efficient magnet-channel interface, low contact resistances and high-spin diffusion length) are achieved in future. The most attractive feature of 3D ASL is evidently the ultra-high area-density. The prospects of achieving ~1000x higher logic density as compared to CMOS may be a motivating factor for the on-going research and experiments in this field.

## 5. Device/Circuit Simulation Framework

In this section we describe the physics based simulation framework used in this work for simulating the ASL multiplier

In order to simulate an ASL device, we need to self-consistently solve both the transport and the magnet dynamics equations. In the four-component spin-circuit model for ASL [5], the channel spin transport is based on the spin diffusion model, The magnet-channel interface is modeled based on the interface model [4]. Both these models are well established and are used for spin transport in long channels [4]. The spin diffusion formulation yields four component conductance matrices $G_{magnet}$, $G_{lead}$, $G_{int}$ and $G_{ch}$ for the elements of nano-magnets, supply leads, magnet-channel interface and the non-magnetic channel, respectively. The four components are the charge and the three spin components. The conductance matrices relate four component voltage drop and current flow between different circuit nodes,

$$[I_c, I_c^z, I_c^x, I_c^y] = [G]_{4\times4}[V_c, V_c^z, V_c^x, V_c^y] \quad (1)$$

The non-magnetic channel and lead elements are modeled as π-conductance matrices with shunt $G_{sh}$ and $G_{se}$ as shunt and series components, respectively [4].

$$G_{sh} = \begin{pmatrix} 0 & 0 & 0 & 0 \\ 0 & g_{sh} & 0 & 0 \\ 0 & 0 & g_{sh} & 0 \\ 0 & 0 & 0 & g_{sh} \end{pmatrix} \quad G_{se} = \begin{pmatrix} \frac{A}{\rho l} & 0 & 0 & 0 \\ 0 & g_{se} & 0 & 0 \\ 0 & 0 & g_{se} & 0 \\ 0 & 0 & 0 & g_{se} \end{pmatrix} \quad (2)$$

Here, $g_{sh} = (A/\rho\lambda)\tanh(l/2\lambda)$ and $g_{se}=(A/\rho\lambda)\text{csch}(l/\lambda)$, l is the length of the contact, A is the area of the contact, ρ is the resistivity and λ is the spin-flip length. These conductance matrices are obtained by solving spin-diffusion equation as shown in [5]. Contact-magnet-channel interface can be described through the matrix $G_{int}$.

$$G_{int} = \begin{pmatrix} g & gP & 0 & 0 \\ gP & g_{se} & 0 & 0 \\ 0 & 0 & \Gamma+\Gamma^* & i(\Gamma-\Gamma^*) \\ 0 & 0 & -i(\Gamma-\Gamma^*) & \Gamma+\Gamma^* \end{pmatrix} \quad (3)$$

where, $g=2-r_lr_l^*-r_rr_r^*$ and $gP=r_rr_r^*-r_lr_l^*$, $\Gamma=1-r_lr_r^*$ and P is the polarization of magnet. $r_l$ and $r_r$ are the reflection coefficients correspond to left and right spin, respectively. The components of the interface matrix are dependent upon the *nano-magnet's* magnetization state, to be evaluated self consistently with magnet dynamics. Note that the elements of $G_{sh}$ are responsible for the decay of spin current along the channel due to spin diffuse scattering [5].

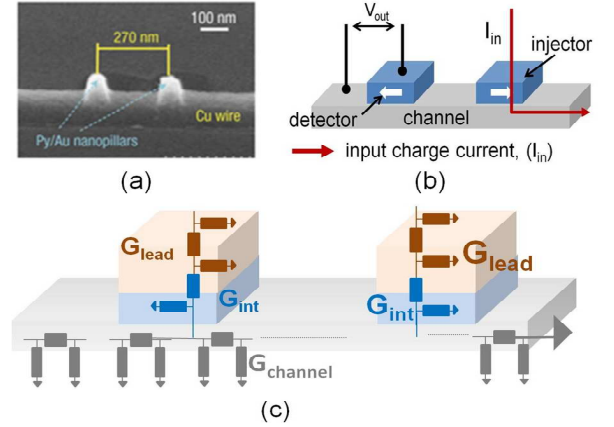

Fig. 19 (a) Fabricated LSV structure in [7], (b) Depiction of structure in fig.1 a, (c) Spin circuit model based on spin diffusion model for the device in fig. 1a

The *Nano-magnet* dynamics is captured by solving the Landau-Lifshitz-Gilbert equation (eq. 4), self-consistently with spin diffusion.

$$\frac{d\hat{m}}{dt} = -|\gamma|\hat{m}\times\vec{H} + \alpha\hat{m}\times\frac{d\hat{m}}{dt} - \frac{1}{qN_s}\hat{m}\times(\hat{m}\times\vec{I}_s) \quad (4)$$

Here m is the magnetization vector, α is the damping constant, $N_S$ is the number of spins in the magnet, γ is gyromagnetic ratio, H is the effective magnetic field and $I_S$ is the spin-current, which is obtained by the transport framework. This simulation-framework has been benchmarked with experimental data on LSV's [1, 3]. This approach leads to the mapping of a spin device structure, involving *nano-magnets* interacting through non-local spin transport, into an equivalent "spin-circuit" [4]. The circuit model for the lateral spin valve is shown in fig. 19. The device parameters used in this work are given in table-I.

TABLE-I
ASL Parameters

| ASL parameters | value |
|---|---|
| Magnet size | 30x15x1 nm$^3$ |
| Ku$_2$V (energy barrier) | 20K$_B$T |
| M$_s$ | 400emu/cm$^3$ |
| α | 0.01 |
| HighP/LowP | 0.9/0.1 |
| Channel Material | Cu/Graphene |
| Spin-flip Length | 1μm/5μm |
| Channel resistivity | 7 ohm-nm |

## 6. CONCLUSION

Although ASL may offer attractive features like scalability and high-density, the energy-efficiency and performance of this logic scheme has been much debated. In this work we explored design-techniques that exploit the specific device-characteristics of ASL to make it comparable to state of the art CMOS technology in terms of energy-efficiency. Leveraging the non-volatility of *nano-magnets* we proposed a 2-phase pipelining scheme that can achieve high-performance for large computing blocks based on ASL. We explored the use of 'leaky' transistors for energy-efficient clocking of ASL that incurs minimal area and power overhead. 3-D ASL design scheme was proposed, which involves stacking of multiple ASL layers that are clocked using the same CMOS transistors. Stacking of ASL layers using the proposed scheme can significantly enhance the power saving as well as area density of ASL through the sharing of CMOS substrate. Simulation results for an FIR filter show that such a design can obtain three orders of magnitude higher density as compared to 15nm CMOS design, while achieving comparable power consumption and performance. The proposed All-Spin-Logic design may therefore be an attractive research path to pursue, to address the possibilities of continuing the Moore's law.

## ACKNOWLEDGEMENT
This work was funded by in part by C-SPIN, National Science Foundation, and Semiconductor Research Corporation.